\RequirePackage{fix-cm}
\documentclass[twocolumn]{svjour3}          
\smartqed  
\usepackage{graphicx}
\usepackage{mathptmx}      
\usepackage{latexsym}
\usepackage[numbers]{natbib}

\usepackage{graphicx}
\usepackage{subfigure}
\usepackage{epsfig}
\usepackage{enumerate}
\usepackage{amsmath}
\usepackage{amssymb}
\usepackage{float}
\usepackage{multirow}
\begin{document}

\title{Machine Learning for Cataract Classification/Grading on Ophthalmic Imaging Modalities: A Survey} 



\author{Xiaoqing Zhang  \and
      Yan Hu \and Zunjie Xiao\and Jiansheng Fang \and Risa Higashita  \and Jiang Liu $^{\ast}$ 
       \thanks{$^{\ast}$ denotes corresponding author}
}

\authorrunning{Short form of author list} 

\institute{XQ Zhang, ZJ Xiao, Y Hu, JS Fang, R Higashita \and J Liu \at Research Institute of Trustworthy Autonomous Systems and Department of Computer Science and Engineering, Southern University of Science and Technology, Shenzhen, China
    \\\email{liuj@sustech.edu.cn}  
}

\date{Received: date / Accepted: date}

\maketitle

\begin{abstract}
	Cataracts are the leading cause of visual impairment and blindness globally. Over the years, researchers have achieved significant progress in developing state-of-the-art machine learning techniques for automatic cataract classification and grading, aiming to prevent cataracts early and improve clinicians' diagnosis efficiency. This survey provides a comprehensive survey of recent advances in machine learning techniques for cataract classification/grading based on ophthalmic images. We summarize existing literature from two research directions: conventional machine learning methods and deep learning methods. This survey also provides insights into existing works of both merits and limitations. In addition, we discuss several challenges of automatic cataract classification/grading based on machine learning techniques and present possible solutions to these challenges for future research.
\end{abstract}

\section{Introduction}
\label{sec:method}

\begin{figure*}
	\centering
	\begin{center}
		\includegraphics[width=0.9\linewidth]{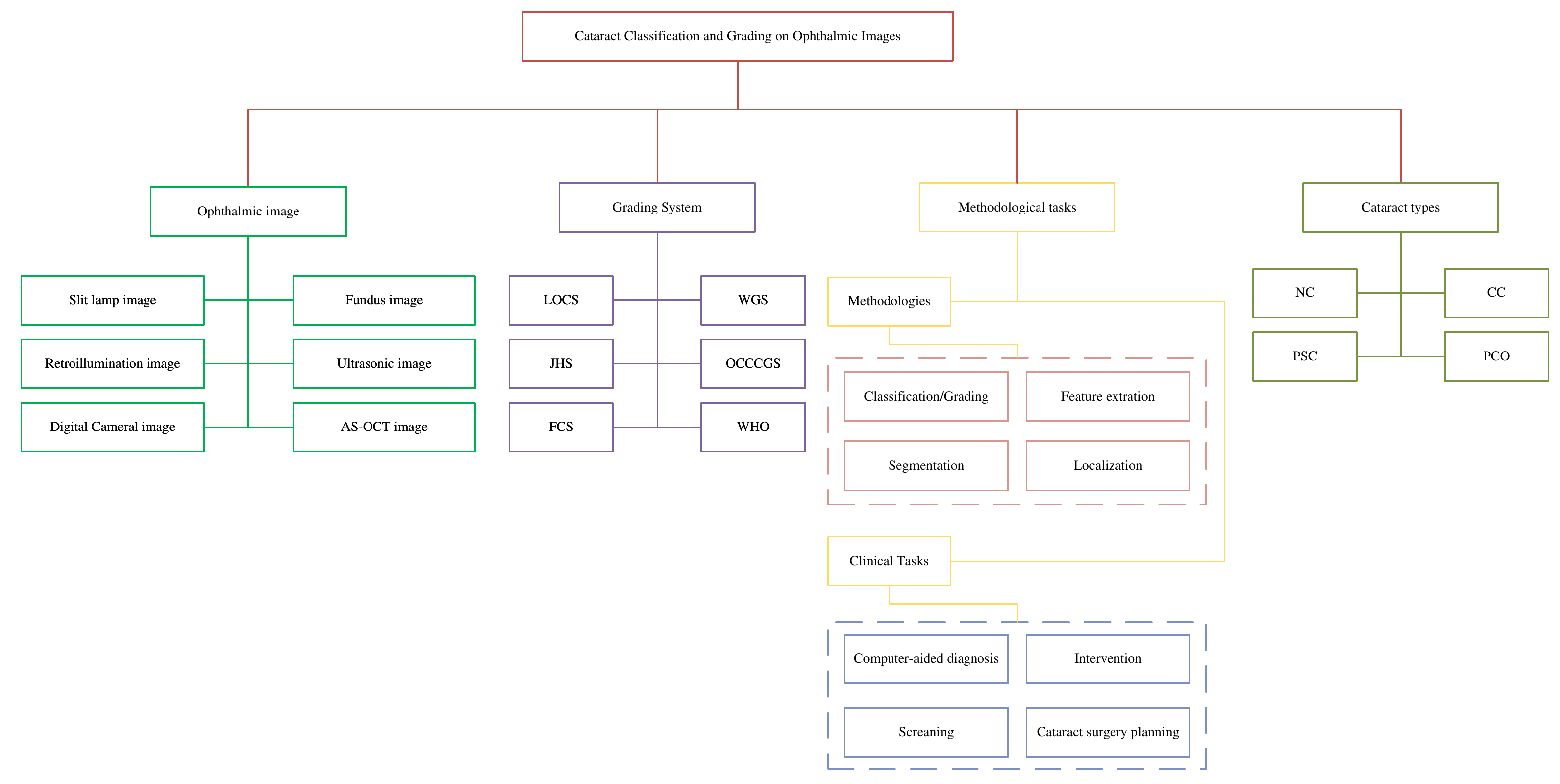}
	\end{center}
	\caption{Overall organization framework of this survey.}
	
	\label{Fig:1}
\end{figure*}

According to World Health Organization (WHO) \cite{bourne2017magnitude,pascolini2012global},
it is estimated that approximately 2.2 billion people suffer visual impairment. Cataract accounts for about 33\% of visual impairment and is the number one cause of blindness (over 50\%) worldwide. Cataract patients can improve life quality and vision through early intervention and cataract surgery, which are efficient methods to reduce blindness ratio and cataract-blindness burden for society simultaneously.

Clinically, cataracts are the loss of crystalline lens transparency, which occur when the protein inside the lens clumps together \cite{asbell2005age}. They are associated with many factors \cite{liu2017cataracts}, such as developmental abnormalities, trauma, metabolic disorders, genetics, drug-induced changes, ages, etc. Genetics and aging are two of the most important factors for cataracts.
According to the causes of cataracts, they can be categorized as age-related cataract, pediatrics cataract (PC), and secondary cataract \cite{asbell2005age, liu2017cataracts}.  
According to the location of the crystalline lens opacity, they can be grouped into nuclear cataract (NC), cortical cataract (CC), and posterior subcapsular cataract (PSC) \cite{li2009automatic, li2010automatic}. NC denotes the gradual clouding and the progressive hardening in the nuclear region. CC is the form of white wedged-shaped and radially oriented opacities, and it develops from the outside edge of the lens toward the center in a spoke-like fashion \cite{liu2017cataracts, chew2012impact}. PSC is granular opacities, and its symptom includes small breadcrumbs or sand particles,  which are sprinkled beneath the lens capsule \cite{li2010automatic}.

Over the past years, ophthalmologists have used several ophthalmic images to diagnose cataract based on their experience and clinical training. This manual diagnosis mode is error-prone, time-consuming, subjective, and costly, which is a great challenge in developing countries or rural communities, where experienced clinicians are scarce. To prevent cataract early and improve the precision and efficiency of cataract diagnosis, researchers have made great efforts in developing computer-aided diagnosis (CAD) techniques for automatic cataract classification/grading \cite{long2017artificial} on different ophthalmic images, including conventional machine learning methods and deep learning methods. The conventional machine learning method is a combination of feature extraction and classification/grading. In the feature extraction stage, a variety of image processing methods have been proposed to obtain visual features of cataract according to different ophthalmic images, such as density-based statistics method, density histogram method, bag-of-features (BOF) method, Gabor Wavelet transform, Gray level Cooccurrence Matrix (GLCM), Haar wavelet transform, etc \cite{huang2009computer, xu2016semantic,li2010automatic, xu2013automatic,Gao2011Computer, srivastava2014automatic, patwari2011detection, fuadah2015performing, pathak2016robust}. In the classification/grading stage, strong classification methods are applied to recognize different cataract severity levels, e.g., support vector machine (SVM) \cite{10.1007/978-3-319-60834-1_34, Yang2015Exploiting,  qiao2017application}. Over the past ten years, deep learning has achieved great success in various fields, including medical image analysis, which can be viewed as a representation learning approach. It can learn low-, mid-, and high-level feature representations from raw data in an end-to-end manner (e.g., ophthalmic images). In the recent, various deep neural networks have been utilized to tackle cataract classification/grading tasks like convolutional neural networks (CNNs), attention-based networks, Faster-RCNN and multilayer perceptron neural networks (MLP). E.g., Zhang et al. \cite{Zhang23} proposed a multi-region fusion attention network to recognize nuclear cataract severity levels.

Previous surveys had summarized cataract types, cataract classification/grading systems, and ophthalmic imaging modalities, respectively \cite{zhang2014survey,liu2017cataracts, shaheen2019survey, lopez2016cataract, zafar2018survey, gali2019cataract,goh2020artificial}; however, none had summarized ML techniques based on ophthalmic imaging modalities for automatic cataract classification/ grading systematically. To the best of our knowledge, this is the first survey that systematically summarizes recent advances in ML techniques for automatic cataract classification/ grading. This survey mainly focuses on surveying ML techniques in cataract classification/grading, comprised of conventional ML methods and deep learning methods. We survey these published papers through Web of Science (WoS), Scopus, and Google Scholar databases. Fig.~\ref{Fig:1} provides a general organization framework for this survey according to collected papers, our summary, and discussion with experienced ophthalmologists.
To understand this survey easily, we also review ophthalmic imaging modalities, cataract grading systems, and commonly-used evaluation measures in brief. Then we introduce ML techniques step by step. We hope this survey can provide a valuable summary of current works and present potential research directions of ML-based cataract classification/grading in the future.

\section{Ophthalmic imaging modalities for cataract classification/grading}
\begin{figure}[htbp]
	\begin{center}
		\includegraphics[width=0.8\linewidth]{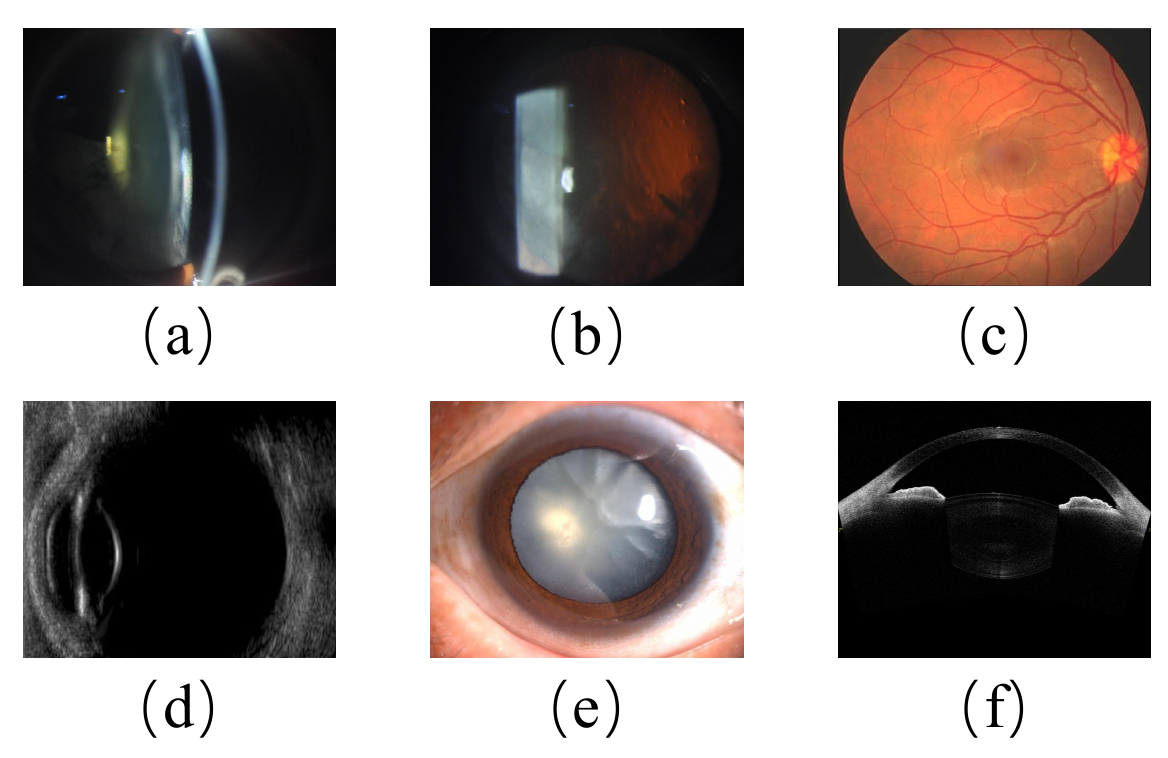}
	\end{center}
	\caption{Six different ophthalmic images. (a) Slit lamp image; (b) Retroillumination image; (c) Ultrasonic image; (d) Fundus image (e) Digital camera image (f) Anterior segment optical
		coherence tomography image.}
	
	\label{fig:2}
\end{figure}

To our best understanding, this survey introduces six different eye images used for cataract classification/grading for the first time: slit-lamp image, retroillumination image, ultrasonic image, fundus image, digital camera image, and anterior segment optical coherence tomography (AS-OCT) image, as shown in Fig.~\ref{fig:2}. In the following section, we will introduce each ophthalmic image type step by step and then discuss their advantages and disadvantages.

\subsection{Slit lamp image}
\begin{figure}[h!]
	\begin{center}
		\includegraphics[width=0.8\linewidth,height =0.2\linewidth]{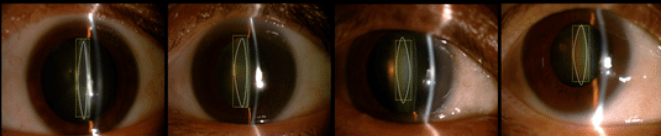}
	\end{center}
	\caption{Slit lamp images with four nuclear cataract severity levels.}
	
	\label{Fig:2}
\end{figure}

The slit lamp camera  \cite{fercher1993slit, waltman1970new} is a high-intensity light source instrument, which is comprised of the corneal microscope and the slit lamp. Silt lamp image can be accessed through slit lamp camera, which is usually used to examine the anterior segment and posterior segment structure of the human eye eyelid, sclera, conjunctiva, iris, crystalline lens, and cornea. Fig.~\ref{Fig:2} offers four representative slit lamp images for four different cataract severity levels.

\subsection{Retroillumination image}
Retroillumination image is a non-stereoscopic medical image, which is accessed through the crystalline lens camera \cite{vivino1995quantitative, gershenzon1999new}. It can be used to diagnose CC and PSC in the crystalline lens region. Two types of retroillumination images through the crystalline lens camera can be obtained: an anterior image focused on the iris, which corresponds to the anterior cortex of the lens, and a posterior image focused on 3-5mm more posteriorly, which intends to image the opacity of PSC. 

\subsection{Ultrasonic image}
In clinical cataract diagnosis, Ultrasound image is a commonly-used ophthalmic image modality to evaluate the hardness of cataract lens objectively \cite{huang2009measurements}. Frequently applied Ultrasound imaging techniques usually are developed based on measuring ultrasonic attenuation and sound speed, which may increase the hardness of the cataract lens \cite{huang2007evaluation}. High-frequency Ultrasound B-mode imaging can be used to monitor local cataract formation, but it cannot measure the lens hardness accurately \cite{tsui2007imaging}. To make up for the B-scan deficiency, the Ultrasound imaging technique built on Nakagami statistical model called Ultrasonic Nakagami imaging \cite{TSUI2010209,tsui2007feasibility,tsui2016acoustic} was developed, which can be used for the visualization of local scatterer concentrations in biological tissues.

\begin{figure}[h!]
	\begin{center}
		\includegraphics[width=0.9\linewidth, height = 2.0cm]{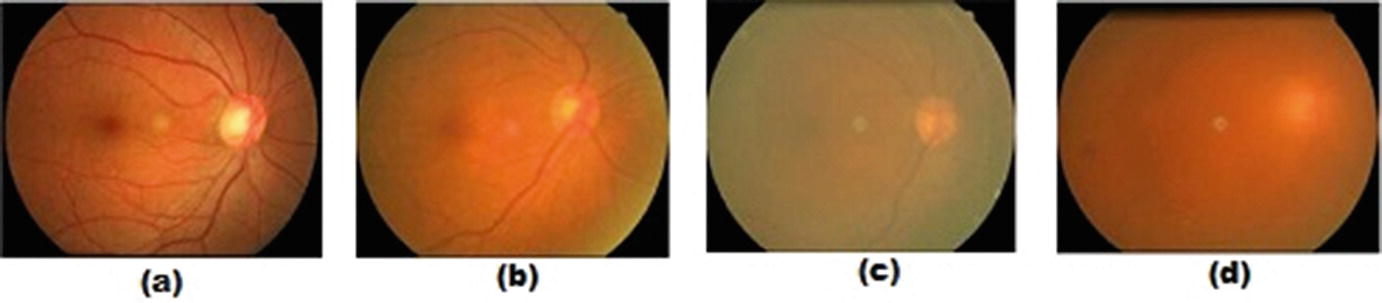}
	\end{center}
	\caption{ Four cataract severity levels on fundus images \cite{cao2020hierarchical}.
		(a) Normal; (b) Immature; (c)
		Mature; (d) Hypermature}
	\label{Fig:3}
\end{figure}

\subsection{Fundus image}
The fundus camera \cite{plesch1987digital, pomerantzeff1979image} is a unique camera in conjunction with a low power microscope, which is usually used to capture fundus images operated by ophthalmologists or professional operators. Fundus image is a highly specialized form of eye imaging and can capture the eye's inner lining and the structures of the back of the eye. Fig.~\ref{Fig:3} shows four fundus images of different cataract severity levels.

\subsection{Digital camera image}
Commonly used digital cameras can access digital camera images like smartphone cameras. Compared with the fundus camera and slit lamp device, the digital camera is easily available and easily used. Hence, using digital cameras for cataract screening has great potential in the future, especially for developing countries and rural areas, where people have limitations to access expensive ophthalmology equipment and experienced ophthalmologists.

\subsection{Anterior segment optical coherence tomography image}
\begin{figure}[!htbp]
	\begin{center}
		\includegraphics[width=0.9\linewidth]{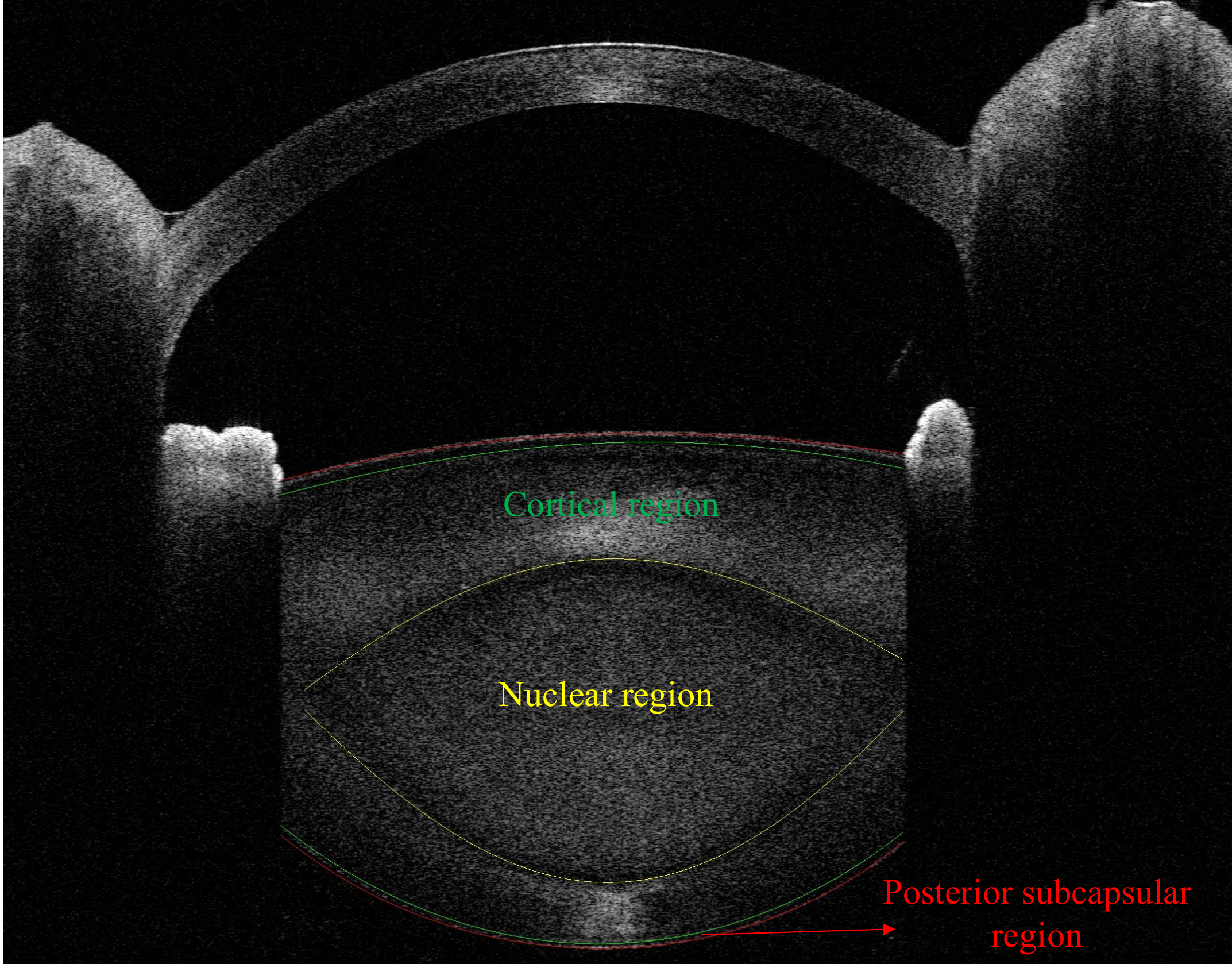}
	\end{center}
	\caption{AS-OCT image. nuclear region is used for NC diagnosis; cortical region is used for CC diagnosis; posterior subcapsular region is used for PSC diagnosis.	
	}
	\label{Fig:4}
\end{figure}

Anterior segment optical coherence tomography (AS-OCT) \cite{ANG2018132} imaging technique is one of optical coherence tomography (OCT) imaging techniques. It can be used to visualize and assess anterior segment ocular features, such as the tear film, cornea, conjunctiva, sclera, rectus muscles, anterior chamber angle structures, and lens \cite{werkmeister2017ultrahigh,  hirnschall2013predicting, yamazaki2014vivo, hirnschall2017prediction}. 
AS-OCT image can provide high-resolution visualization of the crystalline lens in vivo in the eyes of people in real-time without impacting the tissue, which can help ophthalmologists get different information of the crystalline lens through the circumferential scanning mode.
Recent works have suggested that the AS-OCT images can be used to locate the lens region and accurately characterize opacities of different cataract types quantitatively \cite{Grulkowski:18,  pawliczek2020spectral}. Fig.~\ref{Fig:4} offers an AS-OCT image, which can quickly help us know the crystalline lens structure.

\textbf{Discussion:} 
Though six different ophthalmic images are used for cataract diagnosis, slit lamp images and fundus images are the most commonly-used ophthalmic images for clinical cataract diagnosis and scientific research purposes. This is because existing cataract classification/grading systems are built on them. 
Slit lamp images can capture the lens region but cannot distinguish the boundaries between nuclear, cortical, and capsular regions. Hence, it is difficult for clinicians to diagnose different cataract types accurately based on slit lamp images. Fundus images only contain opacity information of cataract and do not contain location information of cataract, which is mainly applied to cataract screening.

Retroillumination images are usually used to diagnose CC and PSC clinically, which have not been widely studied. Digital camera images are ideal ophthalmic images for cataract screening because they can be collected through mobile phones, which are easy and cheap for most people. Like fundus images, digital camera images only have opacity information of cataract but do not contain location information of different cataract types.
Ultrasonic images can capture the lens region and evaluate the hardness of the cataract lens, but they cannot distinguish different sub-regions, e.g., cortex region. AS-OCT image is a new ophthalmic image that can distinguish different sub-regions, e.g., cortex and nucleus regions, significant for cataract surgery planning and cataract diagnosis. However, there is no cataract classification/grading system built on AS-OCTs; thus, it is urgent to develop a clinical cataract classification/grading system based on AS-OCT images. Moreover, existing automatic AS-OCT image-based cataract classification has been rarely studied.

\section{Cataract classification/grading systems}
To classify or grade the severity levels of cataract (lens opacities) accurately and quantitatively, it is crucial and necessary to build standard/gold cataract classification/grading systems for clinical practice and scientific research purposes. This section briefly introduces six existing cataract classification/grading systems.

\subsection{Lens opacity classification system}
\begin{figure}[h!]
	\begin{center}
		\includegraphics[width=0.8\linewidth,height = 0.8\linewidth]{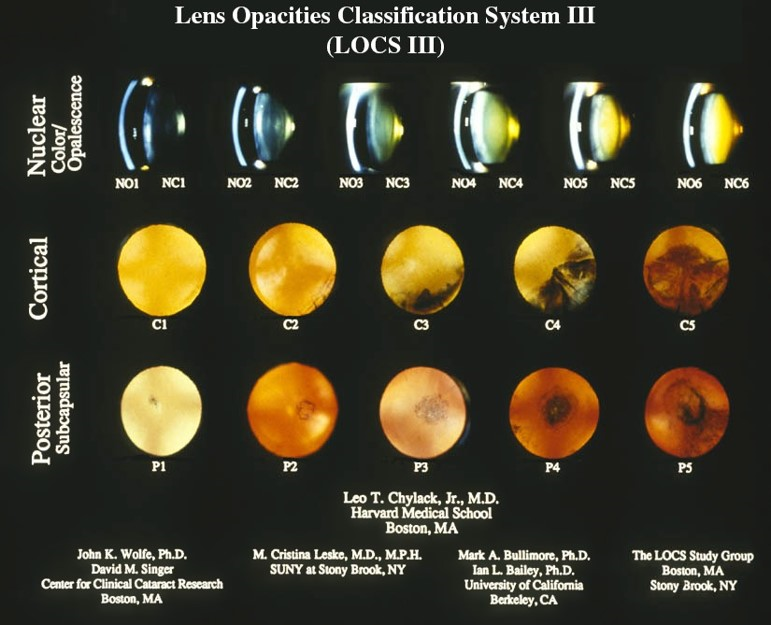}
	\end{center}
	\caption{Lens opacity classification system III.
	}
	\label{Fig:5}
\end{figure}

Lens opacity classification system (LOCS) was first introduced in 1988, which has developed from LOCS I to LOCS III \cite{chylack1993lens,chylack1988lens, chylack1989lens}. LOCS III is widely used for clinical diagnosis and scientific research. In the LOCS III, as shown in Fig.~\ref{Fig:5}, six representative slit lamp images for nuclear cataract grading based on nuclear color and nuclear opalescence; five representative retroillumination images for cortical cataract grading; five representative retroillumination images for grading posterior subcapsular cataract. The cataract severity level is graded on a decimal scale by spacing intervals regularly.

\subsection{Wisconsin grading System}
Wisconsin grading system was proposed by the Wisconsin Survey Research Laboratory in 1990 \cite{klein1990assessment, age2001age, 10.1007/978-3-540-39899-8_73}. It contains four standard photographs for grading cataract severity levels.
The grades for cataract severity levels are as follows: grade 1, as clear or clearer than Standard 1; grade 2, not as clear as Standard 1 but as clear or clearer than Standard 2; grade 3, not as clear as Standard 2 but as clear or clearer than Standard 3; grade 4, not as clear as Standard 3 but as clear or clearer than Standard 4; grade 5, at least as severe as Standard 4; and grade 6, 7 and 8, cannot grade due to severe opacities of the lens (please see detail introduction of Wisconsin grading system in literature \cite{klein1990assessment}). Wisconsin grading system also uses a decimal grade for cataract grading with 0.1-unit interval space, and the range of the decimal grade is from 0.1 to 4.9.

\subsection{Oxford clinical cataract classification and grading system}
Oxford Clinical Cataract Classification and Grading System (OCCCGS) is also a slit-lamp image-based cataract grading system \cite{sparrow1986oxford, hall1997locs}. Different to the LOCS III uses photographic transparencies of the lens as cataract grading standards, it adopts standard diagrams and Munsell color samples to grade the severity of cortical, posterior subcapsular, and nuclear cataract \cite{hall1997locs}. In the OCCCGS, five standard grading levels are used for evaluating the severity level of cataract based on cataract features, such as cortical features, nuclear features, morphological features, etc. \cite{sparrow1986oxford}. E.g., the severity levels of nuclear cataract are graded as follows: Grade 0: No yellow detectable; Grade 1: yellow just detectable; Grade 2: definate yellow; Grade 3: orange yellow; Grade 4: reddish brown; Grade 5: blackish brown \cite{sparrow1986oxford}.

\subsection{Johns Hopkins system}
Johns Hopkins system (JHS) was first proposed in 1980s \cite{west1988use}. It has four standard silt lamp images, which denotes the severity level of cataract based on the opalescence of the lens. For nuclear cataract, Grade 1: opacities that are definitely present but not thought to reduce visual acuity; Grade 2: opacities are consistent with visual acuity between 20/20 and 20/30; Grade 3 opacities are consistent with vision between 20/40 and 20/100; Grade 4: opacities are consistent with the vision of 20/200 or less.

\subsection{WHO cataract grading system}
WHO cataract grading system was developed by a group of experts in WHO \cite{thylefors2002simplified,10665-67221}. The target to develop it is to enable relatively inexperienced observers to grade the most common types of cataracts reliably and efficiently. It uses four severity levels for grading NC, CC, and PSC based on four standard images accordingly.

\subsection{Fundus image-based cataract classification system}
Xu et al. \cite{Xu2010The} proposed a fundus image-based cataract classification system (FCS) through observing the blur level. They used five levels to evaluate the blur levels on fundus images: grade 0: clear; grade 1: the small vessel region was blurred; grade 2: the larger branches of retinal vein or artery were blurred; grade 3: the optic disc region was blurred; grade 4:  The whole fundus image was blurred.

\textbf{Discussion:} From the above-mentioned six existing cataract classification/grading systems, we can conclude that five cataract classification systems are built on slit lamp images, and one is built on fundus images, which can explain that most existing cataract works based on these two ophthalmic imaging modalities. However, these cataract classification/grading systems are subjective due to the limitations of these two imaging devices.
Furthermore, to improve the precision of cataract diagnosis and the efficiency of cataract surgery, it is necessary to develop new and objective cataract classification systems on other ophthalmic image modalities, e.g., AS-OCT images.

\section{Datasets}
In this section, we introduce ophthalmic image datasets used for cataract classification/grading, which can be grouped in private datasets and public datasets.

\subsection{Private datasets}
\textbf{ACHIKO-NC dataset \cite{liu2013integrating}:} 
ACHIKO-NC is the slit-lamp lens images dataset selected from the SiMES I database, used to grade nuclear cataracts. It comprised 5378 images with decimal grading scores (0.3 to 5.0). Professional clinicians determine the grading score of each slit lamp image. ACHIKO-NC is a widely used dataset for automatic nuclear cataract grading according to existing works.

\textbf{ACHIKO-Retro dataset
	\cite{liu2013integrating}:} 
ACHIKO-Retro is the retro-illumination lens image dataset selected from SiMES I database, used to grade CC and PSC. Each lens has two eye image types: anterior image and posterior image. The anterior image focuses on the plane centered in the anterior cortex region, and the posterior image focuses on the posterior capsule region. Most previous CC and PSC grading works were conducted on the ACHIKO-Retro dataset.

\textbf{CC-Cruiser dataset \cite{jiang2021improving}:} CC-Cruiser is the slit lamp image dataset collected from Zhongshan Ophthalmic Center (ZOC) of Sun Yat-Sen University, which is used for cataract screening. It is comprised of 476 normal images and 410 infantile cataract images.

\textbf{Multicenter dataset \cite{jiang2021improving}:} 
Multicenter is the slit lamp image dataset, which is comprised of 336 normal images and 421 infantile cataract images. It was collected from four clinical institutions: the Central Hospital of Wuhan, Shenzhen Eye Hospital, Kaifeng Eye Hospital, and the Second Affiliated Hospital of Fujian Medical University.

\subsection{Public datasets}
EyePACS dataset \cite{cuadros2009eyepacs}:
EyePACS is the fundus image dataset collected from EyePACS, LLC , a free platform for retinopathy screening, used to classify different levels of cataract. It is made available by California Healthcare Foundation. The dataset comprises 88,702 fundus retinal images in which 1000 non-cataract images and 1441 cataract images are provided.

\textbf{HRF dataset \cite{pratap2019computer}:} The high-resolution fundus (HRF) image database is selected from different open-access datasets:
structured analysis of the retina (STARE) \cite{hoover2000locating}, standard diabetic retinopathy database (DIARETDB0) \cite{kauppi2006diaretdb0}, e-ophtha \cite{decenciere2013teleophta}, methods to evaluate segmentation and indexing techniques in the field of retinal ophthalmology (MESSIDOR) database \cite{decenciere2014feedback}, digital retinal images for vessel extraction (DRIVE) database \cite{staal2004ridge}, fundus image registration (FIRE) \cite{hernandez2017fire} dataset, digital retinal images for optic nerve segmentation database (DRIONS-DB) \cite{carmona2008identification}, Indian diabetic retinopathy image dataset (IDRiD) \cite{porwal2018indian}, available datasets released by Dr. Hossein Rabbani \cite{mahmudi2014comparison}, and other Internet sources.

\section{Machine learning techniques}
This section mainly investigates on ML techniques for cataract classification/grading over the years, which is comprised of conventional ML methods and deep learning methods. 

\begin{table*}[h!]
	\footnotesize
	\caption{Conventional ML methods for cataract classification/grading based on ophthalmic images.}
	\centering
	\setlength{\tabcolsep}{0.8mm}{
		\begin{tabular}{ccccccc}
			\hline
			& Literature & Method & Image Type & Year & Application &Cataract Type \\ \hline
			& \cite{li2009automatic} & ASM + SVR  &Slit Lamp Image & 2009 &Grading & NC  \\
			& \cite{huang2009computer}  &ASM + Ranking &Slit Lamp Image &2009 &Grading & NC  \\
			& \cite{li2010computer} & ASM + SVR  &Slit Lamp Image & 2010 &Grading & NC \\  	
			& \cite{huang2010computer}  &ASM + Ranking &Slit Lamp Image & 2010 &Grading & NC \\
			& \cite{li2007towards}   & ASM + LR  &Slit Lamp Image & 2007 &Grading & NC \\
			& \cite{xu2016semantic}  &  SF + SWLR &Slit Lamp Image & 2016 &Grading & NC \\
			& \cite{xu2013automatic}  & BOF + GSR &Slit Lamp Image & 2013 &Grading & NC \\
			& \cite{caixinha2016vivo}  & SVM &Slit Lamp Image & 2016 &Grading & NC\\
			& \cite{Jiang2017Automatic}  & SVM  &Slit Lamp Image & 2017 &Grading & NC  \\
			& \cite{wang2017comparative}  & SVM  &Slit Lamp Image & 2017 &Grading & NC  \\
			& \cite{srivastava2014automatic}  &IGB &Slit Lamp Image &2014 &Grading &NC \\
			& \cite{Jiang2017Automatic}  & RF  &Slit Lamp Image & 2017 &Classification & NC  \\
			& \cite{8400535}  &SRCL &Slit Lamp Image &2018 &Grading &NC \\
			& \cite{jagadale2019computer}  & Hough Circular Transform &Slit Lamp Image&2019 & Classification &NC \\
			& \cite{Kai2019Prediction}  & RF &Slit Lamp Image&2019 & Classification &PC \\
			& \cite{Kai2019Prediction}  & NB&Slit Lamp Image&2019 & Classification &PC \\
			& \cite{li2010automatic}  &Canny + Spatial Filter  &Retroillumination Image &2010 & Classification &PSC \\
			& \cite{li2008automatic}  &EF + PCT  &Retroillumination Image &2008 &Classification &PSC \\
			& \cite{chow2011automatic}  & Canny &Retroillumination Image &2011 &Classification  &PSC \\
			& \cite{ISI:000401634700007}  & MRF &Retroillumination Image &2017 &Classification  &PSC \\
			& \cite{Gao2011Computer}  & LDA &Retroillumination Image &2011 &Classification   &PSC \& CC \\
			& \cite{li2008image}  & Radial-edge \& Region-growing   &Retroillumination Image &2008 &Classification   & CC \\
			& \cite{caxinha2015automatic}  & PCA &Ultrasonic Image &2015  & Classification &Cataract \\
			& \cite{caxinha2015automatic}  & Bayes &Ultrasonic Image &2015  & Classification &Cataract \\
			& \cite{caxinha2015automatic}  & KNN &Ultrasonic Image &2015  & Classification &Cataract \\
			& \cite{caxinha2015automatic}  & SVM &Ultrasonic Image &2015  & Classification &Cataract \\
			& \cite{caxinha2015automatic}  & FLD &Ultrasonic Image &2015  & Classification &Cataract \\
			& \cite{caixinha2016vivo} & SVM &Ultrasonic Image &2016  & Classification &Cataract \\
			& \cite{caixinha2016vivo} & RF &Ultrasonic Image &2016  &  Classification &Cataract \\
			& \cite{caixinha2016vivo} &  Bayes Network &Ultrasonic Image &2016  &  Classification &Cataract \\
			& \cite{caixinha2014new} & PCA + SVM &Ultrasonic Image &2014  &  Classification &Cataract  \\
			& \cite{caixinha2014using}  & Nakagami Distribution +CRT &Ultrasonic Image &2014  &  Classification &Cataract  \\
			& \cite{6518388}  & SVM &Ultrasonic Image &2013  &Classification   &Cataract  \\
			& \cite{fuadah2015performing}  & GLCM + KNN &Digital Camera Image &2015& Classification &Cataract\\
			& \cite{7401368}  & GLCM + KNN &Digital Camera Image &2015& Classification &Cataract\\
			& \cite{pathak2016robust}  &K-Means&Digital Camera Image &2016& Classification &Cataract   \\
			& \cite{patwari2011detection}   &IMF  &Digital Camera Image &2016& Classification &Cataract   \\
			& \cite{10.1007/978-3-319-60834-1_34}   &SVM  &Digital Camera Image &2016& Classification &Cataract   \\
			& \cite{guo2015computer}&WT + MDA  &Fundus Image &2015& Classification&Cataract   \\
			& \cite{Yang2015Exploiting}  &Wavelet-SVM  &Fundus Image &2015& Classification &Cataract   \\
			& \cite{Yang2015Exploiting}  &Texture-SVM  &Fundus Image &2015& Classification &Cataract   \\
			& \cite{Yang2015Exploiting}  &Stacking &Fundus Image &2015& Classification &Cataract   \\	
			& \cite{fan2015principal}  & PCA + Bagging &Fundus Image &2015& Classification &Cataract   \\
			& \cite{fan2015principal}  & PCA + RF &Fundus Image &2015& Classification &Cataract   \\
			& \cite{zhang2019automatic}  & Multi-feature Fusion\&Stacking  &Fundus Image &2019& Classification &Cataract   \\
			& \cite{fan2015principal}  & PCA + GBDT &Fundus Image &2015& Classification &Cataract   \\
			& \cite{fan2015principal} & PCA + SVM &Fundus Image &2015& Classification&Cataract   \\
			& \cite{cao2020hierarchical}  & Haar Wavelet + Voting &Fundus Image &2019& Classification &Cataract   \\
			& \cite{qiao2017application}  & SVM + GA &Fundus Image &2017& Classification &Cataract   \\
			& \cite{huo2019novel}  &AWM + SVM  &Fundus Image &2019& Classification &Cataract   \\
			& \cite{song2016semi} & DT &Fundus Image &2016& Classification &Cataract   \\
			& \cite{song2016semi} & Bayesian Network &Fundus Image &2016& Classification &Cataract   \\
			& \cite{song2019improved} & DWT+SVM &Fundus Image &2019& Classification &Cataract   \\
			& \cite{song2019improved} & SSL &Fundus Image &2019& Classification &Cataract   \\
			& \cite{zhang22} & RF &AS-OCT image &2021& Classification &NC  \\
			& \cite{zhang22} & SVM &AS-OCT image &2021& Classification &NC  \\

			\hline
	\end{tabular}}
	\centering
	\label{Table_1}
\end{table*}

\subsection{Conventional machine learning methods}
\begin{figure}[h!]
	\begin{center}
		\includegraphics[width=0.9\linewidth]{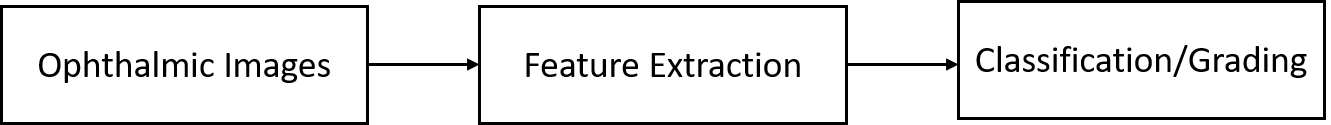}
	\end{center}
	\caption{Flowchart of conventional machine learning based classification and grading.}
	\label{Fig:6}
\end{figure}

Over the past years, scholars have developed massive state-of-the-art conventional ML methods to automatically classify/grade cataract severity levels, aiming to assist clinicians in diagnosing cataract efficiently and accurately. These methods consist of feature extraction and classification /grading, as shown in Fig.~\ref{Fig:6}.
Table~\ref{Table_1} summarizes conventional ML methods for cataract classification/grading based on different ophthalmic images.

\subsubsection{Feature extraction}
Considering the characteristics of different imaging techniques and cataract types, we introduce feature extraction methods based on ophthalmic image modalities.

\textbf{Slit lamp image:} The procedures to extract features from slit lamp images are comprised of the lens structure detection and feature extraction. 

\begin{figure}[h!]
	\begin{center}
		\includegraphics[width=0.9\linewidth, height = 3.0cm]{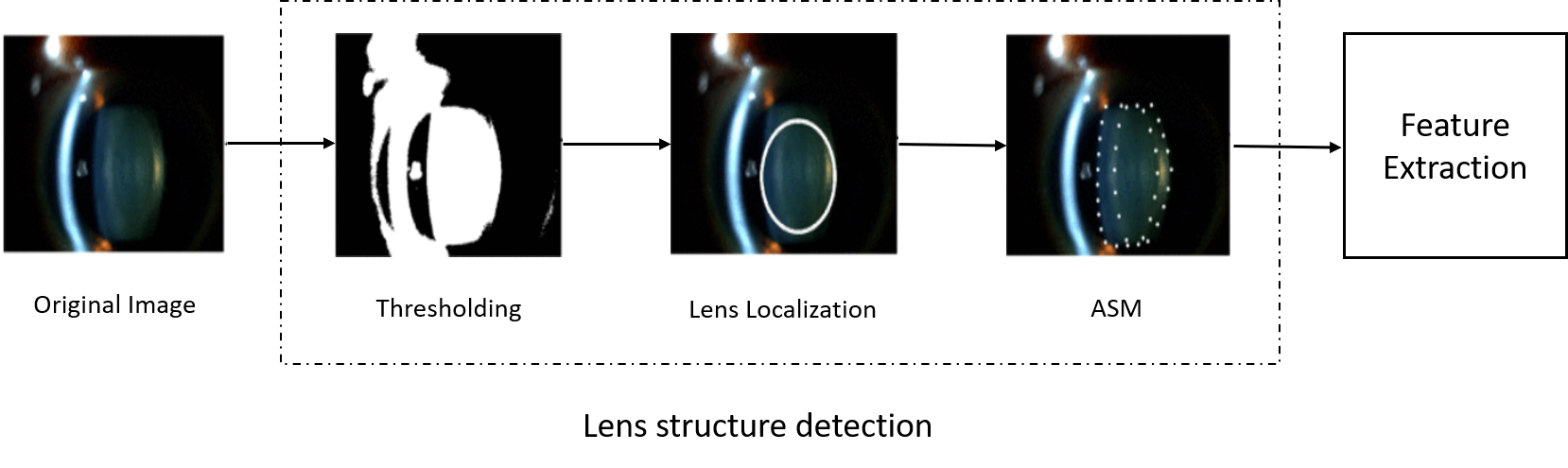}
	\end{center}
	\caption{The procedures to extract image features from slit lamp images for automatic nuclear cataract classification/grading.}
	\label{Fig:7}
\end{figure}
Fig.~\ref{Fig:7} offers a representative slit lamp image-based feature extraction flowchart. Firstly, according to the histogram analysis results of the lens, the foreground of the lens is detected by setting the pixel thresholding, and the background of the lens is even based on slit lamp images. Afterward, we analyze the profile on the horizontal median line of the image. The largest cluster on the line is detected as the lens, and the centroid of the cluster is detected as the horizontal coordinate of the lens center. Then, we get the profile on the vertical line through the point. Finally, the center of the lens is estimated, and the lens can be further estimated as an ellipse with the semi-major axis radius estimated from the horizontal and vertical profile.

The lens contour or shape needs to be captured by following the lens location. Researchers commonly used the active shape model (ASM) method for the lens contour detection \cite{li2010computer, li2010automatic, li2008automatic} and achieved 95\% accuracy of the lens structure detection. The ASM can describe the object shape through an iterative refinement procedure to fit an example of the object into a new image based on the statistical models \cite{li2007towards}. Based on the detected lens contour, many feature extraction methods have been proposed to extract informative features, such as bag-of-features (BOF) method, grading protocol-based method, semantic reconstruction-based method, statistical texture analysis \cite{huang2009computer, xu2016semantic,li2010automatic, xu2013automatic,Gao2011Computer, srivastava2014automatic}, etc.

\begin{figure}[h!]
	\begin{center}
		\includegraphics[width=0.9\linewidth]{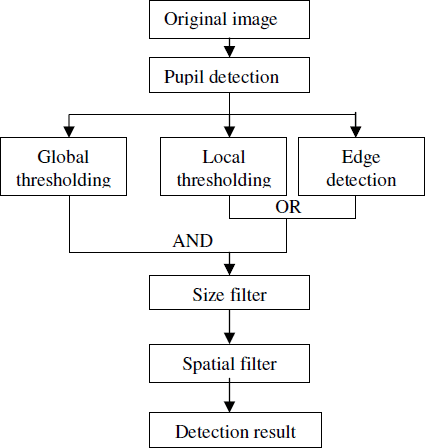}
	\end{center}
	\caption{The procedures to extract features from retroillumination images for cortical cataract (CC) and posterior subcapsular cataract (PSC) diagnosis.}
	\label{Fig:8}
\end{figure}

\textbf{Retroillumination image:} 
It also consists of two stages for feature extraction based on retroillumination images: pupil detection and opacity detection \cite{li2008automatic, li2010automatic, chow2011automatic, Gao2011Computer}, as shown in Fig.~\ref{Fig:8}.
Researchers usually use a combination of the Canny edge detection method, the Laplacian method, and the convex hull method to detect the edge pixel in the pupil detection stage. The non-linear least-square fitting method is used to fit an ellipse based on the detected pixels. In the opacity detection stage, the input image is transformed into the polar coordinate at first. Based on the polar coordinate, classical image processing methods are applied to detect opacity such as global threshold, local threshold, edge detection, and region growing. Apart from the above methods, literature \cite{ ISI:000401634700007} uses Watershed and Markov random fields (MRF) to detect the lens opacity, and results showed that the proposed framework got competitive results of PSC detection.

\textbf{Ultrasound Images \& Digital Camera Images \& AS-OCT Images:} 
For ultrasound images, researchers adopt the Fourier Transform (FT) method, textural analysis method, and probability density to extract features \cite{caixinha2016vivo, caixinha2014new, caixinha2014using}. 

The procedures to extract features from digital camera images is the same to slit lamp images, but different image processing methods are used, such as Gabor Wavelet transform, Gray level Co-occurrence Matrix (GLCM), morphological image feature, and Gaussian filter \cite{  patwari2011detection, fuadah2015performing, pathak2016robust, 7401368}. 

The steps to detect lens region for AS-OCT images are also similar to slit lamp images.
Literature \cite{zhang22} uses intensity-based statistics method and intensity histogram method to extract image features from AS-OCT images.

\begin{figure}[h!]
	\begin{center}		\includegraphics[width=0.9\linewidth, height = 2.5cm]{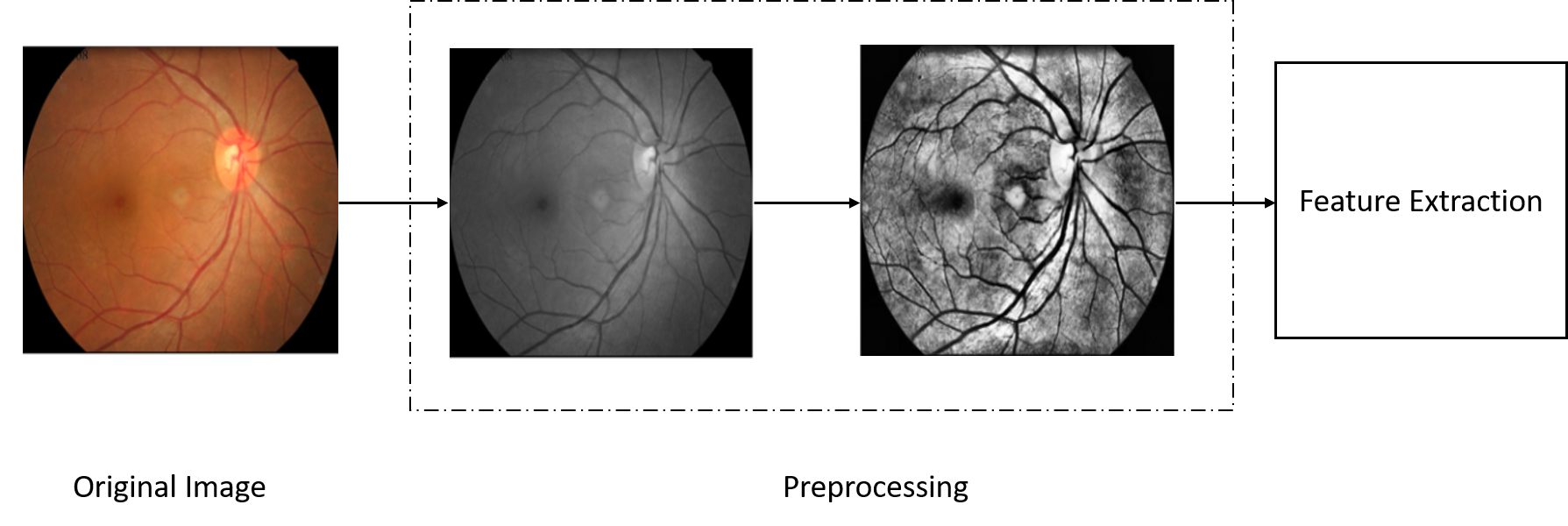}
	\end{center}
	\caption{The procedures to extract features from fundus images for cataract classification or screening.}
	\label{Fig:9}
\end{figure}

\textbf{Fundus image:}
Over the years, researchers have developed various wavelet transform methods to preprocess fundus images for extracting valuable features, as shown in Fig.~\ref{Fig:9}, such as discrete wavelet transform (DWT), discrete cosine transform (DCT), Haar wavelet transform, and top-bottom hat transformation \cite{zhou2019automatic,guo2015computer, cao2020hierarchical}.

\subsubsection{Classification \& grading}
In this section, we mainly introduce conventional machine learning methods for cataract classification/grading.

\textbf{Support vector machine: } Support vector machine (SVM) is a classical supervised machine learning technique, which has been widely be used for classification and linear regression tasks. It is a popular and efficient learning method for medical imaging applications. For the cataract grading task, Li et al. \cite{li2009automatic, li2010computer}  utilized support vector machine regression (SVR) to grade the severity level of cataract and achieved good grading results based on slit lamp images. The SVM classifier is widely used in different ophthalmic images for the cataract classification task. E.g., literature \cite{Jiang2017Automatic, wang2017comparative} applies SVM to classify cataract severity levels based on slit-lamp images. For other ophthalmic image types, SVM also achieves good results based on extracted features \cite{10.1007/978-3-319-60834-1_34, Yang2015Exploiting, Yang2015Exploiting, qiao2017application}.

\textbf{Linear regression:} 
Linear regression (LR) is one of the most well-known ML methods and has been used to address different learning tasks. The concept of LR is still a basis for other advanced techniques, like deep neural networks. Linear functions determine its model in LR, whose parameters are learned from data by training. Literature \cite{li2007towards} first studies automatic cataract grading with LR on slit lamp images and achieves good grading results. Followed by literature \cite{li2007towards}, Xu et al. \cite{ xu2013automatic, xu2016semantic} proposed the group sparsity regression (GSR) and similarity weighted linear reconstruction (SWLR) for cataract grading and achieved better grading results.

\textbf{K-nearest neighbors:} K-nearest neighbors (KNN) method is a simple, easy-to-implement supervised machine learning method used for classification and regression tasks. It uses similarity measures to classify new cases based on stored instances. Y.N. Fuadah et al.\cite{fuadah2015performing} used the KNN to detect cataract on digital camera images and achieved 97.2\% accuracy. Literature \cite{caxinha2015automatic} also uses KNN for cataract classification on Ultrasonic images, which were collected from the animal model.

\textbf{Ensemble learning method: } Ensemble learning method uses multiple machine learning methods to solve the same problem and usually obtain better classification performance. Researchers \cite{Yang2015Exploiting, zhang2019automatic, cao2020hierarchical} have used several ensemble learning methods for cataract classification, such as Stacking, Bagging, and Voting. Ensemble learning methods achieved better cataract grading results than single machine learning methods.

\textbf{Ranking:} Ranking denotes a relationship within the list in a/an descending/ascending order. Researchers \cite{huang2009computer, huang2010computer} applied the ranking strategy to automatic cataract grading by computing the score of each image from the learned ranking function such as RankBoost and Ranking SVM and achieved competitive performance.

\textbf{Other machine learning methods:}
Apart from the above-mentioned conventional ML methods, other advanced ML methods are also proposed for automatic cataract classification/grading, such as Markov random field (MRF), random forest (RF), Bayesian network, linear discriminant analysis (LDA), k-means, and decision tree (DT), etc. E.g., literature \cite{ISI:000401634700007}  applies Markov random field (MRF) to automatic CC classification and achieves good results. \cite{song2019improved} uses semi-supervised learning (SSL) framework for cataract classification based on fundus images and achieves competitive performance.

Furthermore, we can draw the following conclusions:
\begin{itemize} 
	\item For feature extraction, despite the characteristics of ophthalmic images, various image processing techniques are developed to extract useful features, like edge detection method, wavelet transform method, texture extraction methods, etc. However, no previous works systematically compare these feature extraction methods based on the same ophthalmic images, providing a standard benchmark for other researchers. Furthermore, existing works have not verified the effectiveness of a classical feature extraction method on different ophthalmic images, which is significant for the generalization ability of a feature extraction method and building commonly-used feature extraction baselines on cataract classification/grading tasks.
	
	\item For classification/grading, researchers have made great efforts in developing state-of-the-art ML methods in recognizing cataract severity levels on ophthalmic images and demonstrated that ML  methods can achieve competitive performance on extracted features. We found that no existing research has made a comparison between ML methods comprehensively based on the same ophthalmic image or different ophthalmic image types; thus, it is necessary to build conventional ML baselines for cataract classification/grading, which can help researchers reproduce previous works and prompt the development of cataract classification/grading tasks.
	
\end{itemize}

\subsection{Deep learning methods}
In recent years, with the rapid development of deep learning techniques, many deep learning methods ranging from the artificial neural network (ANN), multilayer perceptron (MLP) neural network, backpropagation neural network (BPNN), convolutional neural network (CNN), recurrent neural network (RNN), attention mechanism, to Transformer-based methods, which have been applied to solve different learning tasks such as image classification and medical image segmentation. In this survey, we mainly pay attention to deep learning methods in cataract classification tasks, and Table~\ref{Table_2} provides a  summary of deep learning methods for cataract classification/ grading based on different ophthalmic images.

\begin{table*}
	\caption{Deep learning methods for cataract classification/grading on different ophthalmic images.}
	\centering
	\small
	\begin{tabular}{ccccccc}
		\hline
		&Literature &Method &Image Type &Year &Application &Cataract Type \\ \hline	
		& \cite{gao2015automatic} & CNN+RNN  &Slit Lamp Image & 2015 &Grading & NC  \\
		& \cite{liu2017localization}  & CNN  &Slit Lamp Image & 2017 &Classification & PCO  \\
		& \cite{Jiang2017Automatic}  & CNN  &Slit Lamp Image & 2017 &Classification & PCO  \\
		& \cite{long2017artificial} & CNN  &Slit Lamp Image & 2017 &Classification & Cataract  \\
		& \cite{2018An} & CNN  &Slit Lamp Image & 2018 &Classification & Cataract  \\
		& \cite{jiang2018predicting} & CNN+RNN  &Slit Lamp Image & 2018 &Classification & PCO  \\
		& \cite{xu2019fully} & Faster R-CNN  &Slit Lamp Image & 2019 &Grading & NC  \\
		&\cite{ting2019artificial} & CNN  &Slit Lamp Image & 2019 &Classification & NC  \\
		& \cite{jun2019tournament}  & CNN  &Slit Lamp Image & 2019 &Classification& NC  \\
		&\cite{wu2019universal} & CNN  &Slit Lamp Image & 2019 &Classification & NC  \\
		& \cite{9201392} & CNN  &Slit Lamp Image & 2020 &Classification & NC  \\
		& \cite{jiang2021automatic} & CNN  &Slit Lamp Image & 2021 &Classification & Cataract  \\
		&\cite{jiang2021improving}& R-CNN  &slit lamp Image &2021& Classification & PCO \\
		&\cite{hu2021accv}& CNN  &slit lamp Image &2021& Classification & cataract \\
		& \cite{tawfik2018early}  &  Wavelet + ANN   &Digital Camera Image &2016& Classification&Cataract   \\
		& \cite{caixinha2016vivo} &  MLP &Ultrasonic Image &2016  &  Classification &Cataract \\
		& \cite{9207696} & Attention &Ultrasonic Image &2020  &  Classification &Cataract \\
		& \cite{9533424} & CNN &Ultrasonic Image &2021  &  Classification &Cataract \\
		& \cite{Yang2015Exploiting}  &BPNN  &Fundus Image &2015& Classification &Cataract   \\
		& \cite{8000068} & CNN  &Fundus Image &2017& Classification &Cataract   \\
		& \cite{ dong2017classification} & CNN  &Fundus Image &2017& Classification &Cataract   \\
		& \cite{zhang2019automatic}  & CNN  &Fundus Image &2019& Classification &Cataract   \\
		& \cite{zhou2019automatic}  & MLP &Fundus Image &2019& Classification &Cataract   \\
		& \cite{zhou2019automatic}  & CNN &Fundus Image &2019& Classification &Cataract   \\
		& \cite{zhang2017automatic} & CNN  &Fundus Image &2019& Classification &Cataract   \\
		&\cite{pratap2019computer} & CNN  &Fundus Image &2019& Classification &Cataract   \\
		& \cite{xu2019hybrid} & CNN  &Fundus Image &2020& Classification &Cataract   \\
		& \cite{pratap2021efficient} & CNN  &Fundus Image &2021& Classification &Cataract   \\
		& \cite{imran2021fundus} & CNN+RNN  &Fundus Image &2021& Classification &Cataract   \\
		& \cite{junayed2021cataractnet} & CNN  &Fundus Image &2021& Classification &Cataract   \\
		&\cite{tham2022detecting} & CNN  &Fundus Image &2022& Classification &Cataract   \\
		& \cite{Zhang20NC} & CNN  &AS-OCT Image &2020& Classification & NC \\
		& \cite{xiao20213d} & CNN  &AS-OCT Image &2021& Classification & NC \\
		& \cite{XiaoGCA} & Attention  &AS-OCT Image &2021& Classification & NC \\
		& \cite{Zhang23} & Attention  &AS-OCT Image &2022& Classification & NC \\
		& \cite{HISC-022} & Attention  &AS-OCT Image &2022& Classification & NC \\
		\hline
	\end{tabular}
	\label{Table_2}
\end{table*}


\subsubsection{Multilayer perceptron neural networks}
Multilayer perceptron (MLP) neural network is one type of artificial neural network (ANNs) composed of multiple hidden layers.  Researchers often combined the MLP with hand-engineered feature extraction methods for cataract classification to get expected performance.  Zhou et al. \cite {zhou2019automatic} combined a shallow MLP model with feature extraction methods to classify cataract severity levels automatically. Caixinha et al. \cite{caixinha2014using} utilized the MLP model for the cataract classification and achieved 96.7\% of accuracy. 

Deep neural networks (DNNs) comprise many hidden layers, and they are capable of capturing more informative feature representations from ophthalmic images through comparisons to MLP. DNN usually is used as dense (fully-connected) layers of CNN models. Literature \cite{zhang2017automatic, dong2017classification, zhang2019automatic} uses DNN models for cataract classification and achieves accuracy over 90\% on fundus images.

Recently, several works \cite{tolstikhin2021mlp,lian2021mlp,mansour2021image,touvron2021resmlp,zheng2022mixing} have demonstrated that constructing deep network architectures purely on multi-layer perceptrons (MLPs)  can get competitive performance on ImageNet classification task with spatial and channel-wise token mixing, to aggregate the spatial information and build dependencies among visual features. These MLP-based models have promising results in classical computer vision tasks like image classification, semantic segmentation, and image reconstruction. However, to the best of our knowledge, MLP-based models have not been used to tackle ocular diseases tasks include cataract based on different ophthalmic images, which can be an emerging research direction for cataract classification/grading in the future.


\subsubsection{Convolutional neural networks}
Convolutional neural network (CNN) has been widely used in the ophthalmic image processing field and achieved surprisingly promising performance \cite{szegedy2015going,krizhevsky2012imagenet, chen2015automatic, fu2016deepvessel,abramoff2016improved,litjens2017survey, mansoor2016deep,xie2017aggregated, szegedy2016rethinking, huang2017densely,zhang2018shufflenet}. CNN consists of an input
Layer, multiple convolutional layers, multiple pooling layers, multiple fully-connected layers, and an output layer. The function of convolutional layers is to learn low-, middle, and high-level feature representations from the input images through massive convolution operations in different stages.

For slit lamp images, previous works usually used classical image processing methods to localize the region of interest (ROI) of the lens, and the lens ROI is used as the inputs features of CNN models \cite{liu2017localization, grewal2018deep,   xu2019fully}. E.g., Liu et al.\cite{liu2017localization} proposed a CNN model to detect and grade the severe levels of posterior capsular opacification (PCO), which used the lens ROI as the inputs of CNN. Literature \cite{xu2019fully} uses original images as inputs for CNN models to detect cataract automatically through Faster R-CNN. Literature \cite{long2017artificial} develops an artificial intelligence platform for congenital cataract diagnosis and obtains good diagnosis results. Literature \cite{wu2019universal} proposes a universal artificial intelligence platform for nuclear cataract management and achieves good results. 

For fundus images, literature \cite{xu2019hybrid, zhang2017automatic,dong2017classification, zhang2019automatic} achieves competitive classification results with deep  CNNs. Zhou et al.\cite{zhou2019automatic} proposed the EDST-ResNet model for cataract classification based on fundus images where they used discrete state transition function \cite{deng2018gxnor} as the activation function, to improve the interpretability of CNN models. Fig.~\ref{Fig:11} provides a representative CNN framework for cataract classification on fundus images, which can help audiences know this task easily.

For AS-OCT images, 
literature \cite{yin2018automatic,zhang2019guided, cao2020efficient} proposes CNN-based segmentation frameworks for automatic lens region segmentation based on AS-OCT images, which can help ophthalmologists localize and diagnose different types of cataract efficiently.
Zhang et al. \cite{Zhang20NC} proposed a novel CNN model named GraNet for nuclear cataract classification on AS-OCT images but achieved poor results. \cite{xiao20213d} uses a 3D ResNet architecture for cataract screening based on 3D AS-OCT images. Zhang et al. \cite{HISC-022} tested the NC classification performance of state-of-the-art CNNs like ResNet, VGG, and GoogleNet on AS-OCT images, and the results showed that EfficientNet achieved the best performance.

\begin{figure}[h!]
	\begin{center}		\includegraphics[width=0.9\linewidth,height=4.0cm]{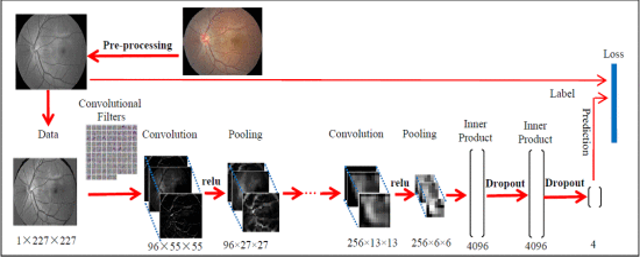}
	\end{center}
	\caption{ An example of a convolutional neural network (CNN) model for cataract classification on fundus image \cite{xu2019hybrid}.}
	\label{Fig:12}
\end{figure}

\subsubsection{Recurrent neural networks}

Recurrent neural network (RNN) is a typical feedforward neural network architecture where connections between nodes form a directed or undirected graph along a temporal sequence. RNN are skilled at processing sequential data effectively for various learning tasks \cite{hu2020harmonic}, e.g., speech recognition. Over the past decades, many RNN variants have been developed to address different tasks, where Long short-term memory (LSTM) network is the most representative one. However, researchers have not used pure RNN architecture to classify cataract severity levels yet, which can be a research direction for automatic cataract classification.

\subsubsection{Attention mechanisms}
\begin{figure}[h!]
	\begin{center}\includegraphics[width=0.9\linewidth, height=0.6\linewidth]{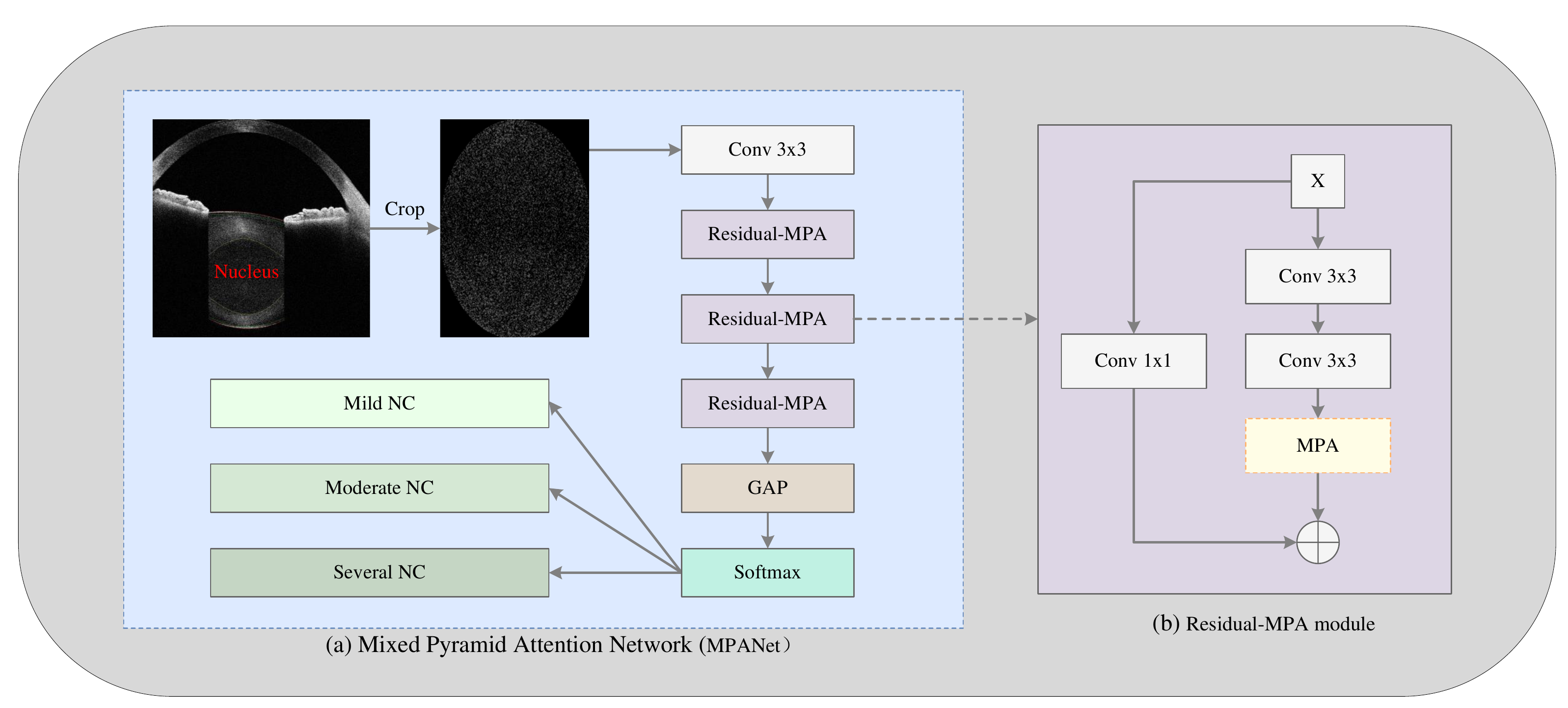}
	\end{center}
	\caption{ An example of attention-based CNN architecture for cataract classification on AS-OCT images.}
	\label{Fig:11}
\end{figure}

Over the past years, attention mechanisms have been proven useful in various fields, such as computer vision \cite{choi2020channel}, natural language processing (NLP) \cite{vaswani2017attention}, and medical data processing \cite{zhang2019net}. Generally, attention mechanism can be taken an adaptive weighting process in a local-global manner according to feature representations of feature maps. In computer vision, attention can be classified into five categories: channel, spatial, temporal attention, branch attention, and attention combinations such as channel \& spatial attention \cite{guo2021attention}. Each attention category has a different influence on the computer vision field. Researchers have recently used attention-based CNN models for cataract classification on different ophthalmic images. Zhang et al. \cite{9207696}  proposed a residual attention-based network for cataract detection on Ultrasound Images. Xiao et al. \cite{XiaoGCA} applied a gated attention network (GCA-Net) to classify nuclear cataract severity levels on AS-OCT images and got good performance. \cite{HISC-022} presents a mixed pyramid attention network for AS-OCT image-based nuclear cataract classification in which they construct the mixed pyramid attention (MPA) block by considering the relative significance of local-global feature representations and different feature representation types, as shown in Fig.~\ref{Fig:11}.

Especially, self-attention is one representative kind of attention mechanism. Due to its effectiveness in capturing capture long-range dependencies and generality, it has been playing an increasingly important role in a variety of learning tasks \cite{vaswani2017attention,wang2018non,yu2021metaformer}. Massive deep self-attention networks (e.g., Transformer) have achieved state-of-the-art performance through mainstream CNNs on visual tasks. Vision Transformer (ViT) \cite{dosovitskiy2020image} is a first-pure transformer architecture proposed for image classification and gets promising performance. Recently, researchers also extended the transformer-based models for different medical image analysis tasks \cite{he2022transformers}; however, no current research work has utilized transformer-based architectures to recognize cataract severity levels.

\subsubsection{Hybrid neural networks}
TA hybrid neural network indicates that a neural network is comprised of two or more two deep neural network types. In recent years, researchers have increasingly used hybrid neural networks to address different learning tasks \cite{gulati2020conformer}. Due to its ability in inheriting the advantages from different neural network architectures, such as CNN, MLP, and transformer. Literature \cite{jiang2018predicting, gao2015automatic,imran2021fundus} proposes the hybrid neural network models for cataract classification /grading by considering the characteristics of RNN and CNN models, respectively. In the future, we believe that more advanced hybrid neural network models will be designed for cataract classification/grading based on different ophthalmic image modalities.

\begin{figure}[h!]
	\begin{center}		\includegraphics[width=0.9\linewidth,height=0.4\linewidth]{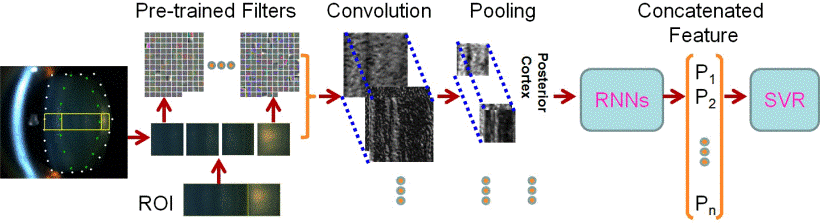}
	\end{center}
	\caption{A hybrid neural network model for nuclear cataract grading on slit lamp images, comprised of a CNN model and a RNN model \cite{gao2015automatic}.}
	\label{Fig:13}
\end{figure}

\textbf{Discussion:}
From Table~\ref{Table_1} and Table~\ref{Table_2}, we can conclude as follows:

\begin{itemize}
	\item  \textbf{Ophthalmic image perspective:} Slit lamp images and fundus images account for most automatic cataract classification/grading works, this is because these two ophthalmic image modalities are easy to access and have the clinical gold standards. Except for ultrasound images collected from the animal model and human subjects, five other ophthalmic imaging modalities were used for automatic cataract classification/grading.
	
	\item \textbf{Cataract classification/grading perspective:} Existing works mainly focused on cataract screening, and most of them achieved over 80\% accuracy. The number of cataract classification works is more than the number of cataract grading works. Since discrete labels are easy to access and be confirmed through comparisons continuous labels.
	
	\item \textbf{Publication year perspective:}  Conventional machine learning methods were first used to classify or grade cataract automatically. With the emergence of deep learning, researchers have gradually constructed advanced deep neural network architectures to automatically predict cataract severity levels due to their surprising performance. However, the interpretability of deep learning methods is worse than conventional machine learning methods.
	
	\item \textbf{Learning paradigm perspective:} Most previous machine learning methods belong to supervised learning methods, and only two existing methods used semi-supervised learning methods. Specifically, unsupervised and semi-supervised learning methods have achieved competitive performance in computer vision and NLP, which have not been widely applied to automatic cataract diagnosis.
	
	\item \textbf{Deep neural network architecture perspective:} According to Table~\ref{Table_2}, we can see that CNNs account for over 50\% deep neural network architectures. Two reasons can explain that: 1) Ophthalmic images are the most commonly-used way for cataract diagnosis by clinicians. 2) Compared with RNN and MLP, CNN are skilled at processing image data. To enhance precision of cataract diagnosis, it is better to combine image data with non-imaging data, e.g., age.
	
	\item \textbf{Performance comparison on private/public datasets:} Pratap et al. \cite{pratap2021efficient} made a comparison of AlexNet, GoogleNet, ResNet, ResNet50, and SqueezeNet for the cataract classification on the public EyePACS dataset, and results showed all CNN models obtained over 86\% in the accuracy, and AlexNet got the best performance. \cite{HISC-022} offers the NC classification results of attention-based CNNs (CBAM, ECA, GCA, SPA, and SE on the private NC dataset, and results showed SE obtained better performance than other strong attention methods. 
	Furthermore, \cite{xu2019hybrid,xu2019fully, pratap2021efficient, pratap2019computer} used pre-trained CNNs for automatic cataract classification on the EyePACS dataset according to transfer learning strategy, and they also concluded that training pre-trained CNNs benefited the cataract classification performance than training CNNs from scratch.
	
	\item \textbf{Data augmentation techniques:} Researchers use the commonly-used method for cataract classification/grading to augment ophthalmic image data for deep neural networks, such as flipping, cropping, rotation, and noise injection. To further validate or enhance the generalization ability of deep neural network models on cataract classification/grading tasks, other data augmentation techniques should be considered: translation, color space transformations, random erasing, adversarial training, meta-learning, etc \cite{shorten2019survey}.
\end{itemize}

\section{Evaluation measures}
This section introduces evaluation measures to assess the performance of cataract classification/grading. In this survey, Classification denotes the cataract labels used for learning are discrete, e.g., 1,2,3,4, while grading denotes cataract labels are continuous, such as 0.1,0.5,1.0, 1.6, and 3.3.

For cataract classification, accuracy (ACC), sensitivity (recall), specificity, precision, F1-measure (F1), and G-mean are commonly used to evaluate the classification performance. \cite{guo2015computer, cao2020hierarchical,jiang2018predicting}.

\begin{equation}
ACC=\frac{TP + TN}{TP + TN + FP + FN},
\label{eq1}
\end{equation}
\begin{equation}
Sensitivity=\frac{TP}{TP + FN},
\label{eq2}
\end{equation}

\begin{equation}
Specificity = \frac{TN}{TN + FP},
\label{eq3}
\end{equation}

\begin{equation}
Precision = \frac{TP}{TP + FP},
\label{eq4}
\end{equation}

\begin{equation}
F1 = \frac{2 * precision * recall}{precision + recall},
\label{eq5}
\end{equation}

\begin{equation}
G-mean = \sqrt{
	\frac{TP}{TP+FN} * \frac{TN}{TN+FP}},
\label{eq6}
\end{equation}
where TP, FP, TN, and FN denote the numbers of true positives, false positives, true negatives, and
false negatives, respectively. Other evaluation measures like receiver operating characteristic curve (ROC), the area under the ROC curve (AUC), and kappa coefficient \cite{cao2020hierarchical} are also applied to measure the overall performance.

For cataract grading, the following evaluation measures are used to evaluate the overall performance, titled the exact integral agreement ratio $R_0$,  the percentage of decimal
grading errors $\leqslant  0.5$ $R_{e0.5}$, the percentage of decimal grading errors $\leqslant 1.0$ $R_{e1.0}$, and the mean absolute error $\varepsilon$ \cite{li2007towards, cheung2011validity, xu2016semantic, 4582838, huang2010computer}. 

\begin{equation}
R_{0} = \frac{\lvert \left \lceil G_{gt} \right \rceil  = \left \lceil G_{pr}  \right \rceil \rvert_{0}}{N},
\label{eq7}
\end{equation}

\begin{equation}
R_{e0.5} = \frac{ \lvert \lvert G_{gt}- G_{pr}\rvert\leqslant 0.5 \rvert_{0}}{N},
\label{eq8}
\end{equation}

\begin{equation}
R_{e1.0} = \frac{ \lvert \lvert G_{gt}- G_{pr}\rvert\leqslant 1.0 \rvert_{0}}{N},
\label{eq9}
\end{equation}

\begin{equation}
\varepsilon =\frac{\sum \lvert G_{gt}-G_{pr} \rvert}{N},
\label{eq10}
\end{equation}
where $G_{gt}$ and $G_{pr}$ denote the ground-truth grade and the predicted
grade. $\left \lceil . \right \rceil $ is the ceiling function, $\lvert. \rvert $ is the absolute function, $\lvert . \rvert_{0}$ a function
that counts the number of non-zero values, and N denotes the number of
images.

\section{Challenges and possible solutions}
Although researchers have made significant development in automatic cataract classification/grading over the years, this field still has challenges. This section presents these challenges and gives possible solutions.

\subsection{Lacking public cataract datasets}
PPublic cataract datasets are a very critical issue for cataract classification/grading. Previous works have made tremendous progress in automatic cataract/grading \cite{xu2013automatic, doi:10.1080/09286580600878844,  caixinha2014new, guo2015computer}, there is no public and standard ophthalmology image dataset available except for public challenges datasets and multiple ocular diseases. Hence, it is difficult for researchers to follow previous works because the cataract dataset is unavailable.
To this problem, it is necessary and significant to build public and standard ophthalmology image datasets based on standardized medical data collection and storage protocol. Public cataract datasets can be collected from hospitals and clinics with ophthalmologists' help. This dataset collection mode can ensure the quality and diversity of cataract data and help researchers develop more advanced ML methods.

\subsection{\textbf{Developing standard cataract classification/grading protocols based on new ophthalmic imaging modalities}}  

Most existing works used the LOCS III as the clinical gold diagnosis standard to grade/classify cataract severity levels for scientific research purposes and clinical practice. However, the LOCS III is developed based on slit-lamp images, which may not work well for ophthalmic images, such as fundus images, AS-OCT images, and Ultrasonic images. To solve this problem, researchers have made much effort in constructing new cataract classification/grading standards for other ophthalmic image types. E.g., researchers refer to the WHO Cataract Grading System and develop a cataract classification/grading protocol for fundus images \cite{Xu2010The} based on clinical research and practice, which is widely used in automatic fundus-based cataract classification \cite{xu2019hybrid, zhang2017automatic,dong2017classification, zhang2019automatic}.

To address the issue of developing standard cataract classification/grading protocols for new eye images, e.g., AS-OCT images. In this survey, we propose two possible solutions for reference as follows.

\begin{itemize}
	\item Developing a cataract grading protocol based on the clinical verification. Literature \cite{Grulkowski:18,  pawliczek2020spectral} uses AS-OCT images to observe the lens opacities of patients based on the LOCS III in clinical practice, and statistical results showed that high correlation between cataract severity levels and clinical features with inter-class and intra-class analysis. The clinical finding may provide clinical support and reference. Hence, it is likely to develop a referenced cataract classification/grading protocol for ASOCT images based on the clinical verification method like the LOCS III.	
	\item Building the mapping relationship between two ophthalmic imaging modalities. The lens opacity is an important indicator to measure the severity level of cataracts, which are presented on different ophthalmic images through different forms in clinical research. Therefore, it is potential to construct the mapping relationship between two different ophthalmic imaging modalities to develop a new cataract classification/grading protocol by comparing the lens opacities, e.g., fundus image-based cataract classification system WHO Cataract Grading System.
\end{itemize}

Furthermore, to verify the effectiveness of new standard cataract grading protocols, we must collect multiple center data from hospitals in different regions.

\subsection{\textbf{How to annotate cataract images accurately}} 
Data annotation is a challenging problem for the medical image analysis field including cataract image analysis, since it is the significant base for accurate ML-based disease diagnosis. However, clinicians cannot label massive cataract images manually \cite{Resnikoff588,resnikoff2012number}, because it is expensive, time-consuming, and objective. To address this challenges, we offer the following solutions:

\begin{itemize}
	\item \textbf{Semi-supervised learning:}
	\cite{ song2016semi,song2019improved} uses the semi-supervised learning strategy to recognize cataract severity levels on fundus images and achieves expected performance.	It is probably to utilize weakly supervised learning methods to learn useful information from labeled cataract images and let the method automatically label unlabeled cataract images according to learned information. 
	
	\item \textbf{Unsupervised learning:}
	Recent works have shown that deep clustering/unsupervised learning techniques can help researchers acquire labels positively rather than acquire labels negatively \cite{ravi2016optimization, grira2004unsupervised, fogel2019clustering, guo2017deep, hsu2015neural}. Thus, We can actively apply deep clustering/unsupervised learning methods to label cataract images in the future. 
	
	\item \textbf{Content-based image retrieval:}
	the content-based image retrieval (CBIR) \cite{ramamurthy2011content, sivakamasundari2014proposal, fathabad2012content,9313536} technique has been widely used for different tasks based on different image features, which also can be utilized to annotate cataract images by comparing testing images with standard images.
\end{itemize}

\subsection{How to classify/grade cataract accurately for precise cataract diagnosis}

Most previous works focused on cataract screening, and few works considered clinical cataract diagnosis, especially for cataract surgery planning. This is because different cataract severity levels and cataract types should clinically take the corresponding treatments.  Hence, it is necessary to develop state-of-the-art methods to classify cataract severity levels accurately, and this survey provides the following research directions.

\begin{itemize}
	\item \textbf{Clinical prior knowledge injection:}  Furthermore, cataracts are associated with various factors \cite{lin2020practical}, e.g., sub-lens regions   \cite{xu2016semantic,Gao2011Computer,caxinha2015automatic, Grulkowski:18,  pawliczek2020spectral}, which can be considered as domain knowledge of cataract. Thus, we can infuse the domain knowledge into deep networks for automatic cataract classification/grading according to the characteristics of ophthalmic images. E.g., \cite{HISC-022} incorporates clinical features in attention-based network design for classification.
	
	\item \textbf{Multi-task learning for classification and segmentation:} Over the past decades, multi-task learning techniques have been successfully applied to various fields, including medical image analysis. Xu et al. \cite{xu2019fully} used the Faster-RCNN framework to detect the lens region and grade nuclear cataract severity levels on slit lamp images and achieved competitive performance. Literature \cite{yin2018automatic,zhang2019guided} proposes the deep segmentation network framework for automatic lens subregion segmentation based on AS-OCT images, which is a significant base for cataract diagnosis and cataract surgery planning. Moreover, more multi-task learning frameworks should be developed for cataract classification and lens segmentation, considering multi-task learning framework usually keeps a good balance between performance and complexity.
	
	\item \textbf{Transfer learning:} Over the years, researchers have used the transfer learning method to improve the cataract classification performance \cite{xu2019hybrid,xu2019fully} with pre-trained CNNs. Large deep neural network models usually perform better than small deep neural network models based on massive data, which previous works have demonstrated. 
	However, it is challenging to collect massive data in the medical field; thus, it is vital for us to develop transfer learning strategies to take full use of pre-trained parameters for large deep neural network models, to further improve the performance of cataract-related tasks.
	
	\item \textbf{Multimodality learning:} Previous works only used ophthalmic image type for cataract diagnosis \cite{xu2016semantic, Gao2011Computer,caxinha2015automatic, yang2013classification}, multimodality learning \cite{cardoso2017deep, article,8599078, cheng2015multimodal} have been utilized to tackle different medical image tasks, which also can be used for automatic cataract classification/grading based on multi-ophthalmic images or the combination of ophthalmic images and non-image data.  Multimodality data can be classified into image data and non-image data. However, it is challenging to use multimodality images for automatic cataract classification. Two reasons can account for it: 1) only silt lamp images and fundus images have standard cataract classification systems, which have no correlation relationship between them; 2) existing classification systems are subjective, it is challenging for clinicians to label two cataract severity levels correctly for different ophthalmic images. Furthermore, we can combine image data and non-image data for automatic cataract diagnosis because it is easy to collect non-image data such as age and sex associated with cataracts. Furthermore, recent studies have \cite{mohammadi2012using,lin2020practical} used non-image data for PC and PCO diagnosis, which demonstrated that it is potential to use multimodality data to improve the precision of cataract diagnosis
	
	\item \textbf{Image denoising:} Image noise is an important factor in affecting automatic cataract diagnosis on ophthalmic images. Researchers have proposed different methods to remove the noise from the images based on the characteristics of ophthalmic images, such as Gaussian filter, discrete wavelet transform (DWT), discrete cosine transform (DCT), Haar wavelet transform, and its variants \cite{zhou2019automatic,guo2015computer, 7401368}. Additionally, recent research has begun to use the GAN model for medical image denoising and achieved good results.
\end{itemize}

\subsection{Improving the interpretability of deep learning methods} 
Deep learning methods have been widely used for cataract classification/grading. However, deep learning methods are still considered a black box without explaining why they make good prediction results or poor prediction results. Therefore, it is necessary to give reliable explanations for the predicted results based on deep learning methods. Literature \cite{zhou2019automatic} visualizes weight distributions of deep convolutional layers and dense layers to explain the predicted results of deep learning methods. It is likely to analyze the similarities between the weights of convolutional layers to describe the predicted results.

\cite{zhang2019guided} indicated that infusing the domain knowledge into networks can improve the cataract segmentation results based on AS-OCT images and enhance the interpretability. 
It is promising for researchers to combine deep learning methods with different domain knowledge to enhance the interpretability of deep networks.

Modern neural networks \cite{guo2017calibration} usually are poorly calibrated, \cite{muller2019does} indicated that improving the confidence of predicted probability can explain the predicted results. It is a potential solution for researchers to enhance interpretability by enhancing the confidence of predicted probability. 

Moreover, researchers have achieved great progress in the interpretability of deep learning methods in the computer vision field.\cite{zhou2016learning, selvaraju2017grad, guidotti2018survey, lu2017knowing} used the heat maps to localize discriminative regions where CNNs learned.
Hinton et al. \cite{2019Similarity} proposed the correlation analysis technique to analyze similarities of neural network representations to explain what neural networks had learned. These interpretability techniques are also used for deep learning-based cataract classification works for improving the interpretability.

\subsection{Moblie cataract screening}
In the past years, researchers have used digital camera images for cataract screening based on ML techniques and have achieved good results \cite{supriyanti2011achievement, patwari2011detection, fuadah2015performing, supriyanti2012consideration, pathak2016robust, rana2017cataract, 7401368}. Digital camera images can be accessed through mobile phones, which are widely used worldwide. Thus, there are endless potentials to embed powerful digital cameras into mobile phones for cataract screening. Although deep learning methods can achieve excellent cataract screening results on digital camera images, their parameters are too much. One possible solution is to develop lightweight network models. Researchers have achieved significant improvement on designing lightweight deep learning methods over the years \cite{cheng2017survey, jiang2019model, he2017channel}, which demonstrated that deep networks kept a good balance between accuracy and speed, which are important for developing cataract screening-based lightweight deep networks.

\subsection{How to evaluate the generalization ability of machine learning methods for other eye disease classification tasks}

Although researchers have proposed massive machine learning techniques for automatic cataract classification/grading, including conventional machine learning methods and deep learning methods, and obtained promising performance on different eye images. However, only Zhang et al. \cite{HISC-022} validated the generalization ability of the proposed attention-based network on cataract and age-related macular degeneration (AMD) classification task, while others focused on the pure cataract classification task. The following reasons can explain that: 1) due to characteristics of eye images, it is difficult to validate the effectiveness of a method on different eye disease classification tasks, especially for conventional machine learning methods. 2) most existing works focused on specific eye disease tasks like cataract and glaucoma rather than universal methods for medical analysis. 3) Appropriate open-access medical datasets are rare. 4) None previous works have systematically concluded the existing methods and built standard baselines.

This survey offers the following solutions:
\begin{itemize}
	\item More open-access medical datasets are supposed to be released.
	\item it is significant to build baselines of different eye disease tasks, which can provide the standard benchmarks for others to follow. 
\end{itemize}

\subsection{Others}

Apart from the above-mentioned challenges, there are other challenges for medical image analysis, including cataract, such as label noises \cite{song2022learning}, domain distribution shift across different datasets, and how to train more robust models. However, these challenges have not been studied in the cataract classification/grading field. Considering the real diagnosis requirements, it is necessary to address these challenges by following advanced machine learning methods, such as graph convolution method  \cite{zhong2019graph}, knowledge distillation, adversarial learning, etc.

\section{Conclusion}
Automatic and objective cataract classification/grading is an important and hot research topic, helping reduce blindness ratio and improve life quality. In this survey, we survey the recent advances in the ML-based cataract classification/grading on ophthalmic imaging modalities, and highlight two research fields: conventional ML and deep learning methods. 
Though ML techniques have made significant progress in cataract classification and grading, there is still room for improvement. First, public cataract ophthalmic images are needed, which can help build ML baselines. Secondly, automatic ophthalmic image annotation has not been studied seriously and is a potential research direction. Thirdly, developing accurate and objective ML methods for clinical cataract diagnosis is efficient to help more people improve their vision. It is necessary to enhance the interpretability of deep learning, which can prompt applications in cataract diagnosis. Finally, previous works have used digital camera images for cataract screening, which is a good idea by devising lightweight ML techniques that address the cataract screening problem for developing countries with limited access to expensive ophthalmic devices.


\section*{Abbreviation}
CAD: computer aided diagnosis; ML: machine learning; WHO: World Health Organization; NC: nuclear cataract; CC: cortical cataract; PSC: posterior subcapsular cataract; LR: linear regression ; RF: random forest; SVR: support vector machine regression; ASM: active shape model; NLP: natural language
processing; OCT:  optical coherence
tomography; LOCS: Lens Opacity Classification System; OCCCGS: Oxford Clinical Cataract Classification and Grading
System; SVM: support vector
machine; artificial intelligent: AI; BOF: bag-of-features; PCA: principal
component analysis; DWT: discrete wavelet transform; DCT: discrete cosine transform; KNN: K-nearest neighbor; ANN: artificial neural network; BPNN: back propagation neural network; CNN: convolutional neural network; RNN: recurrent neural network; DNN: deep neural network; MLP: multilayer perceptron; ROI: region of interes; AE: autoencoder; GAN: generative adversarial network; CBIR: content-based image retrieval; ASOCT: anterior segment ptical coherence tomography; 3D: three-dimensional; 2D: two-dimensional; SSL:Semi-supervised learning; AWM: Adaptive window model; GSR: Group Sparsity Regression; GLCM: gray level Co-occurrence matrix;
Image morphological feature: IMF; SWLR: Similarity Weighted Linear Reconstruction; CBAM: Convolutional block attention module; ECA: Efficient channel attention; SE: squeeze-and-excitation; SPA: spatial pyramid attention.

\bibliographystyle{spbasic}      
\bibliography{mir}       

%
%

\end{document}